\documentclass[epj]{webofc}
\usepackage[varg]{txfonts}   % Web of Conferences font
\usepackage{units}
\wocname{EPJ Web of Conferences}
\woctitle{ARENA 2016}
\begin{document}
\title{Recent results from the ARIANNA neutrino experiment}

\author{\firstname{Anna} \lastname{Nelles}\inst{1}\fnsep\thanks{\email{anelles@uci.edu}}
        \firstname{for the} \lastname{ARIANNA Collaboration}
        % etc.
}

\institute{Department of Physics and Astronomy, University of California Irvine, 92697 CA, USA }

\abstract{%
 The ARIANNA experiment is currently taking data in its pilot-phase on the Ross ice-shelf. Fully autonomous stations measure radio signals in the frequency range from 100 MHz to 1 GHz. The seven station HRA was completed in December 2014, and augmented by two special purpose stations with unique configurations. In its full extent ARIANNA is targeted at detecting interactions of cosmogenic neutrinos ($>10^{16}$eV) in the ice-shelf. Downward-pointing antennas installed at the surface will record the radio emission created by neutrino-induced showers in the ice and exploit the fact that the ice-water surface acts as a mirror for radio emission. ARIANNA stations are independent, low-powered, easy to install and equipped with real-time communication via satellite modems. We report on the current status of the HRA, as well as air shower detections that have been made over the past year. Furthermore, we will discuss the search for neutrino emission, future plans and the energy-dependent sensitivity of the experiment.
}
\maketitle
\section{Introduction}
\begin{figure*}
\centering
\includegraphics[width=0.25\textwidth]{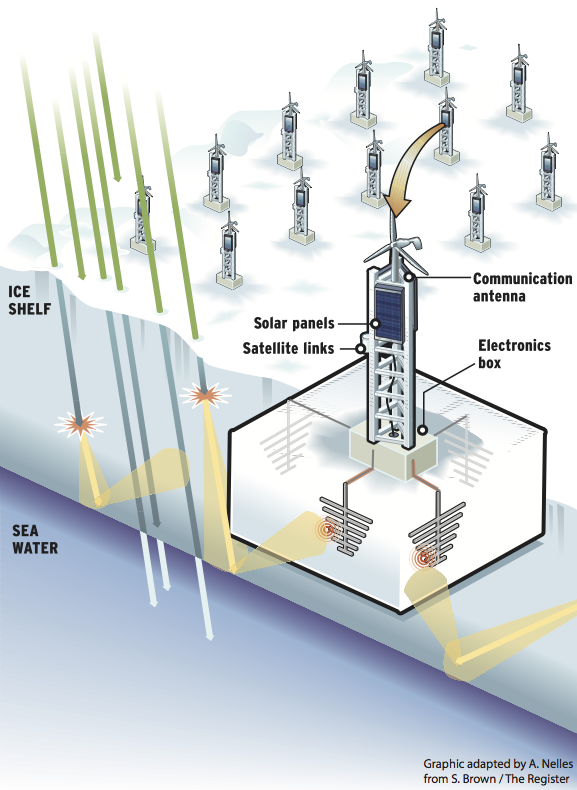} \hspace{0.3cm}
\includegraphics[width=0.7\textwidth]{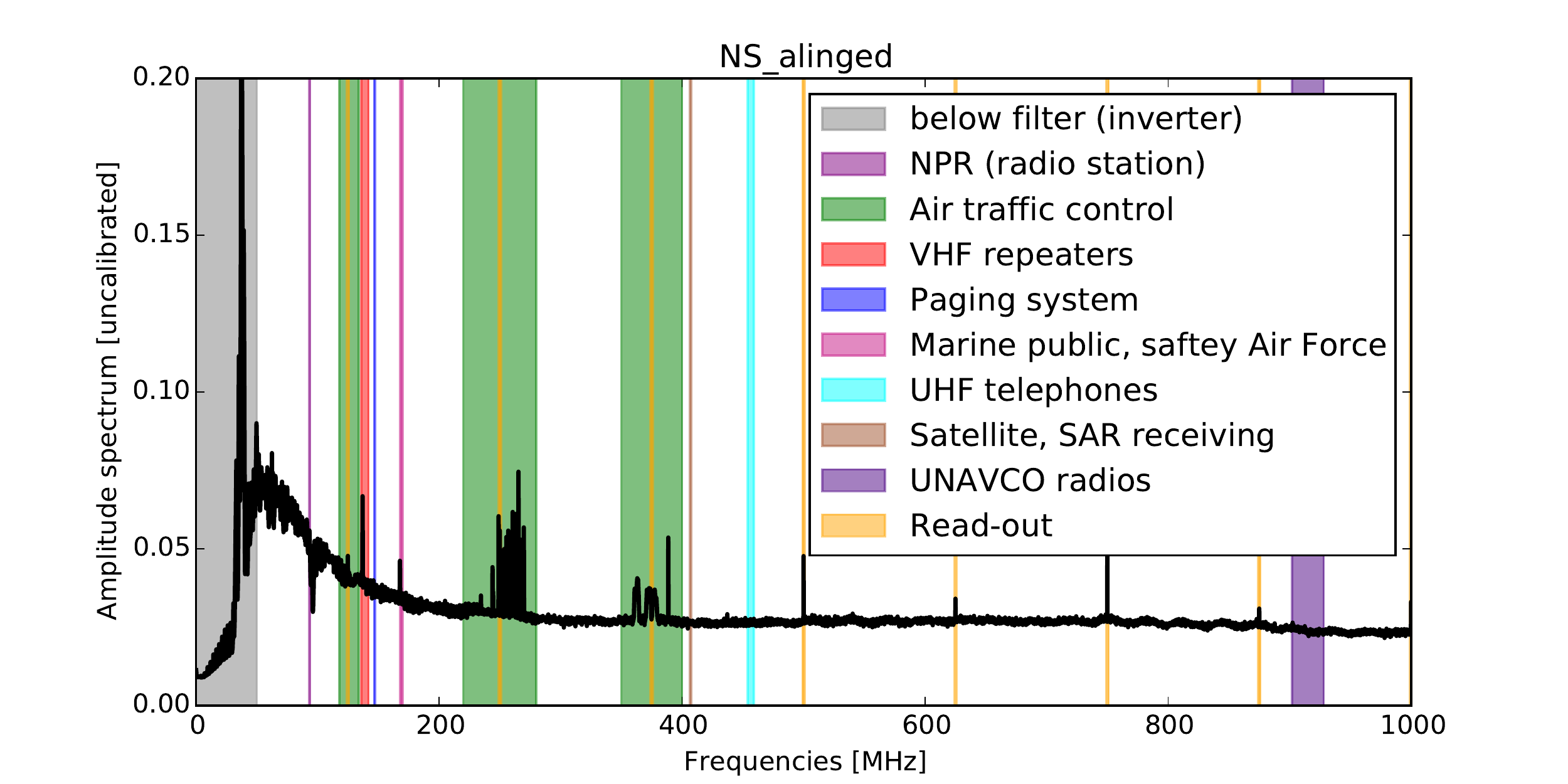}
\caption{On the left side an artist's impression of the ARIANNA array and the detection concept is shown. On the right measurements of the background noise down to \unit[50]{MHz} are shown. This data has been averaged over several measurements to reveal the narrow-band transmitters.}
\label{fig:ARIANNA}       % Give a unique label
\end{figure*}

The ARIANNA detector is aimed at detecting neutrinos at an energy above \unit[$10^{16}$]{eV} \cite{AstroP}. Whenever these neutrinos produce an electromagnetic cascade after interacting in the ice of Moore's Bay on the Ross Ice-shelf in Antarctica, the ensuing radio emission will be strong enough to be detectable in ARIANNA. The detector profits from the fact that the water-ice interface at the bottom of the shelf-ice acts as a mirror for the radio emission and the antennas can be installed on top of the ice (see Figure \ref{fig:ARIANNA}). Since the antenna geometry is not defined by deep holes in the ice, high-gain antennas can be installed that cover the frequency range from 50 to 1000 MHz. 

\section{Current status of the HRA}

\begin{figure*}
\centering
\includegraphics[width=0.3\textwidth]{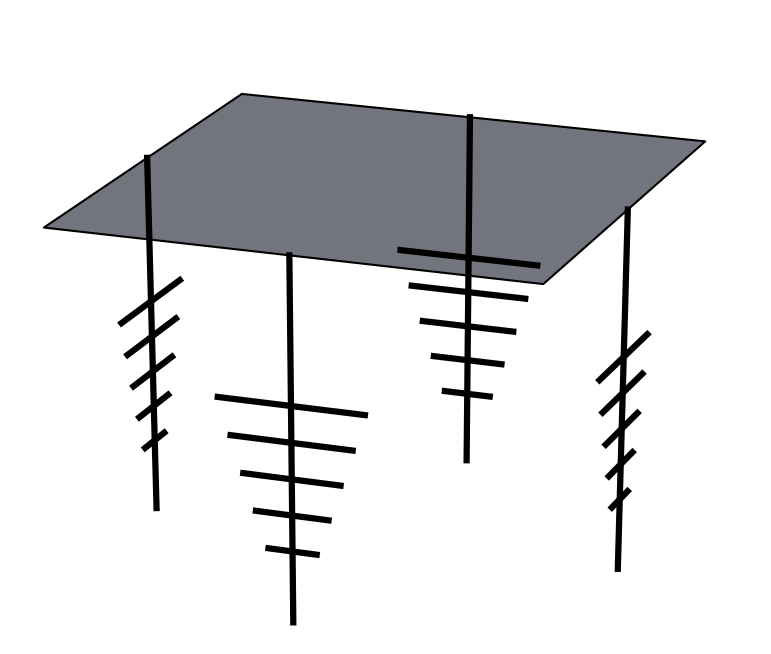}
\includegraphics[width=0.3\textwidth]{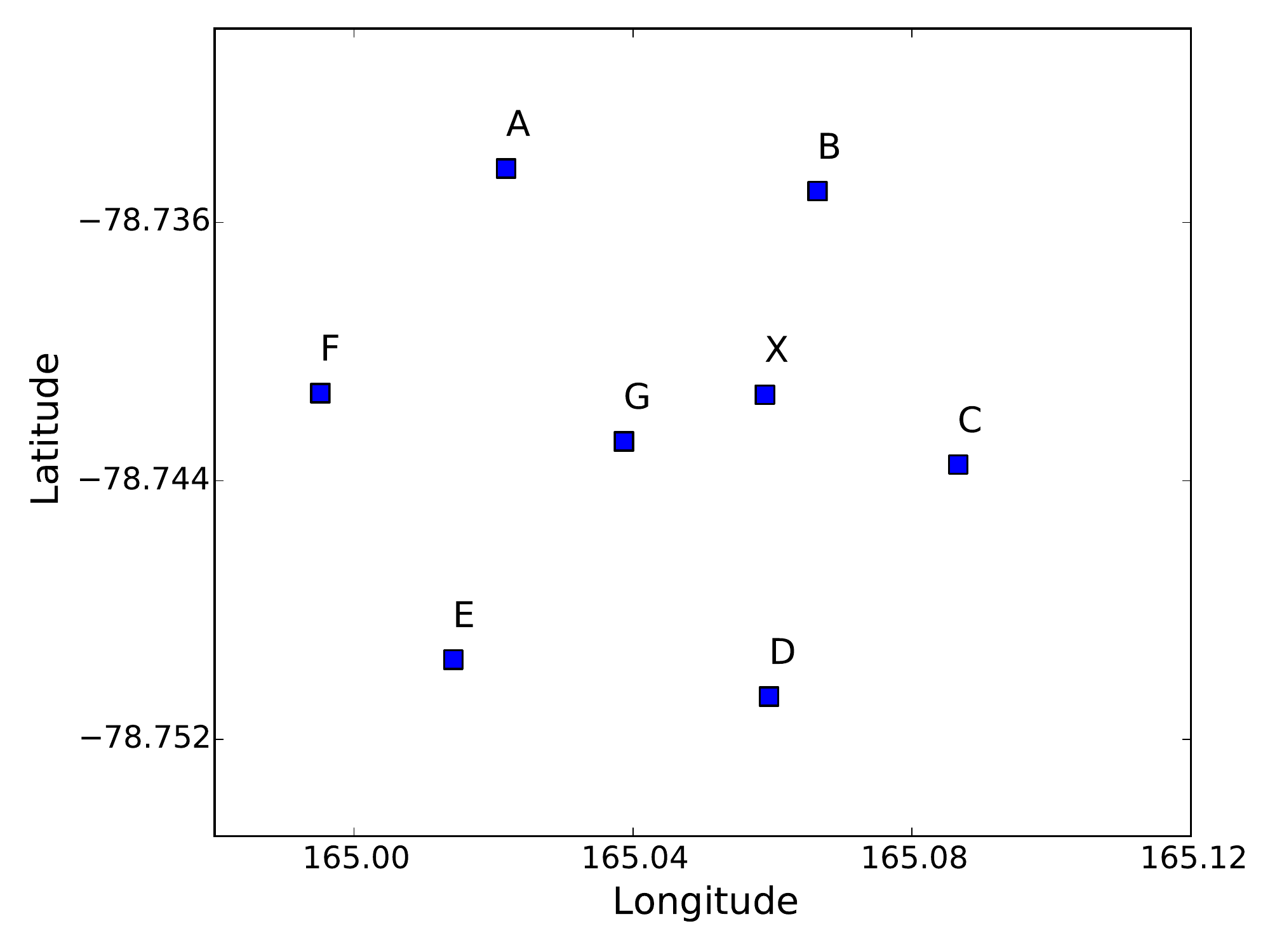}
\includegraphics[width=0.3\textwidth]{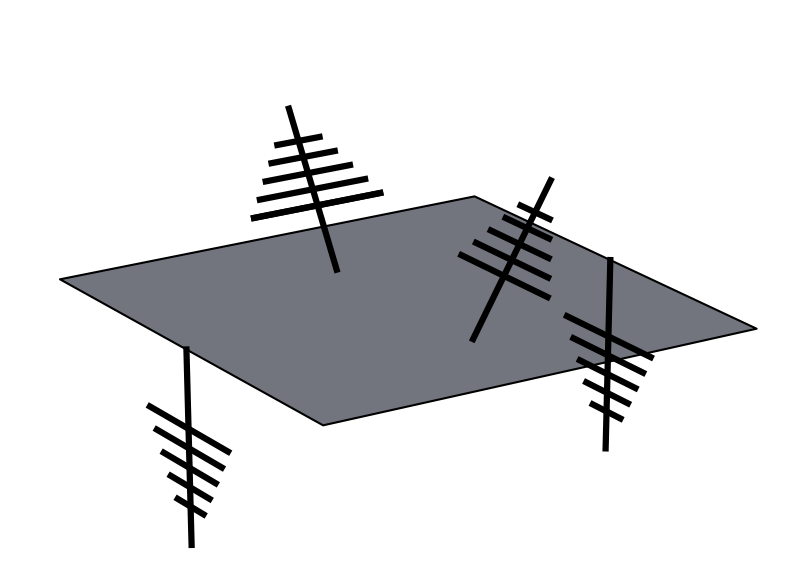}
\caption{The HRA detector layout. At positions A - G (middle) stations with four antennas pointing down towards the water-ice interface are installed (left). At position X a station with two upward facing antennas is installed (right), which has been targeted at air shower detection.}
\label{fig:ARIANNA-layout}       % Give a unique label
\end{figure*}

The ARIANNA detector is currently in its pilot phase, the hexagonal-radio array (HRA) \cite{IEEE}. The main array consists of 7 stations, each with four down-ward facing antennas as shown on the left in Figure \ref{fig:ARIANNA-layout}. In addition, there is one station where two antennas have been rotated upwards as shown on the right in the same figure. While the regular stations are prototypes for the neutrino capability of ARIANNA, the altered station is a test set-up for the rejection of air shower signals. It has been found in simulations that the up-to-down ratio in amplitude is fully sufficient to veto air shower signals coming from above when looking for neutrino signals coming from below. 
All HRA stations are equipped with solar panels and battery systems and run autonomously during the daylight-season. The stations communicate either via long-range wifi or via satellite communication. Both allow for real-time data-transfer off Antarctica. 
The stations are operated with custom digitizers and amplifiers and are optimized for frequencies between 100 and 1000 MHz, with increased gain below 500 MHz. The data is sampled at 1 GHz on one board for the four antennas per stations, which delivers very high relative timing precision in the ps range. The ARIANNA Collaboration is currently developing an eight-channel alternative, which will allow for the full combination of down-and upward facing antennas in one station. The stations usually trigger on a coincidence of two or more antennas. A trigger in a station is formed, if a combination of low-and-high threshold crossings is found in a window of \unit[5]{ns}. This allows for the HRA stations to run on very low threshold values of between three and four times the noise level and still maintain a very low trigger rate. 

\begin{figure*}
\centering
\includegraphics[width=0.7\textwidth]{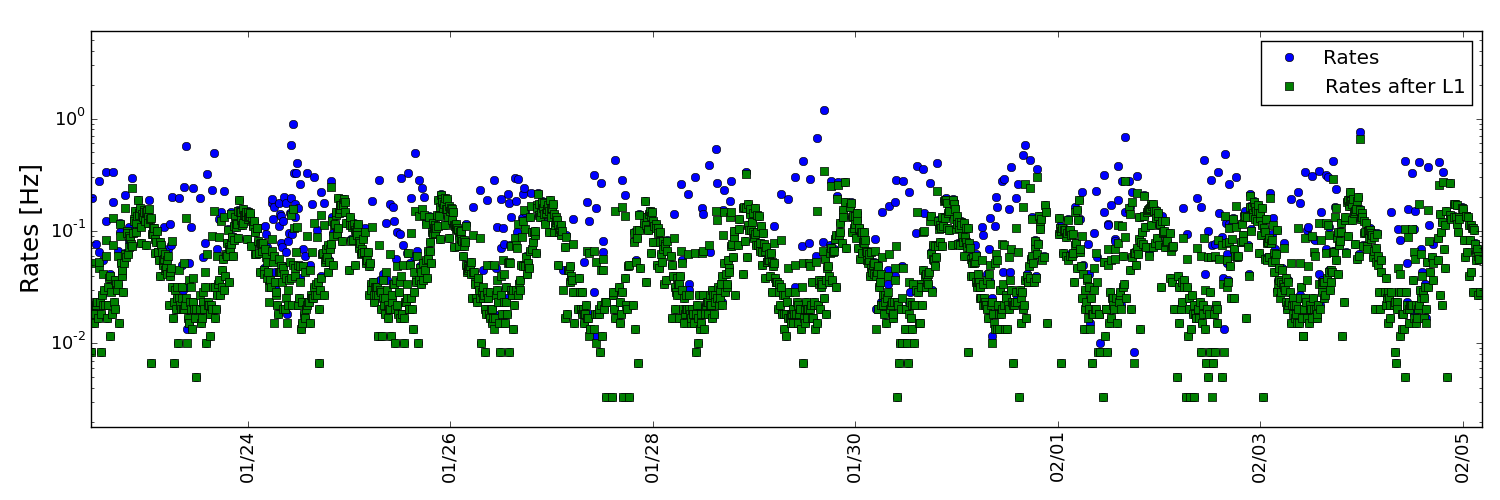}
\caption{Trigger rates as function of time in station X triggering on the upward facing antennas. The blue circles indicate the raw trigger rates, before the cut on narrowband transmitters (L1, see text) has been applied. The green squares are the thus reduced trigger rates. The oscillation in the trigger rates has a period of 23 hours and 56 minutes. It is therefore linked to an astronomical phenomenon. }
\label{fig:rates_galaxy}       % Give a unique label
\end{figure*}

\begin{figure*}
\centering
\includegraphics[width=0.7\textwidth]{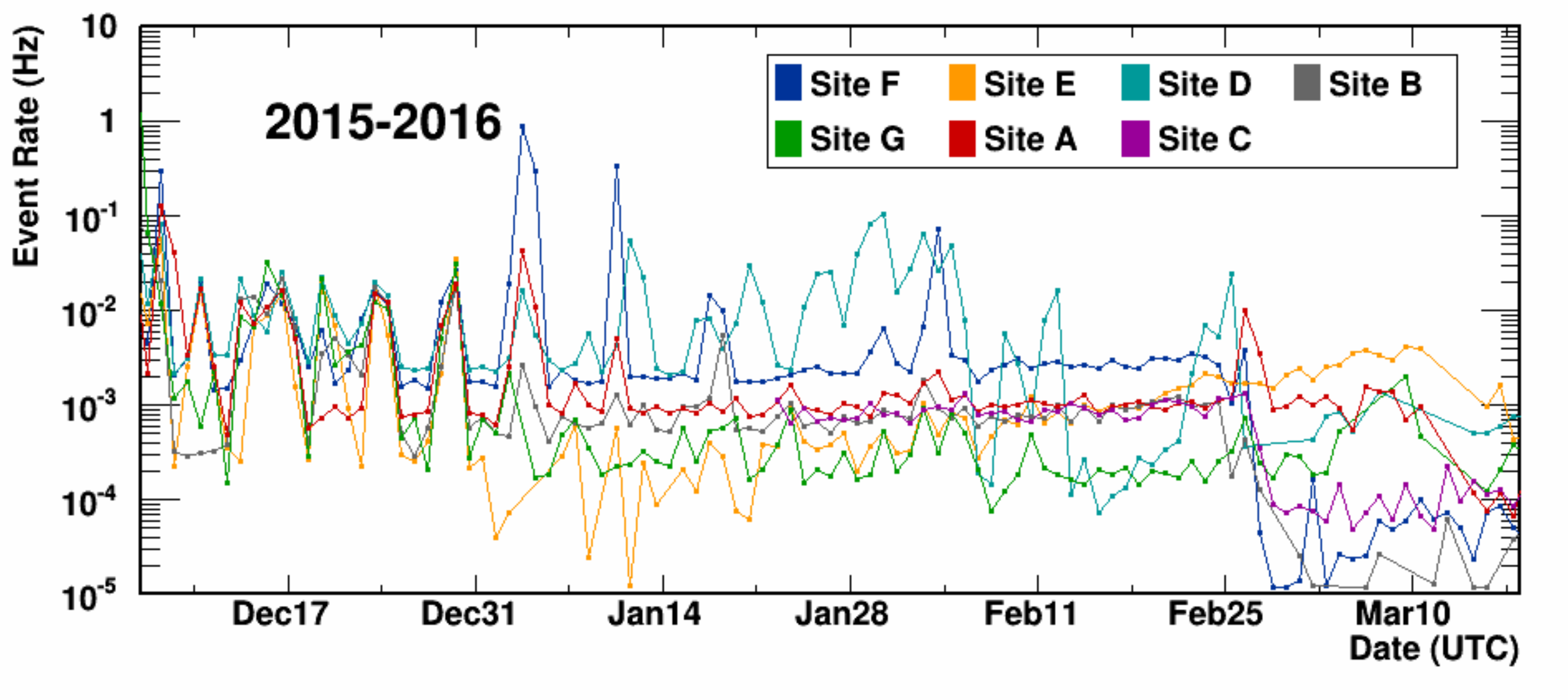}
\caption{Trigger rates as function of time of all regular HRA stations. After December 31st a veto for narrowband transmitters (L1) has been introduced gradually. After February 25th the thresholds have been raised to lower the data rate to allow for data transfer only via satellite modem and turning off the long-distance wifi. Also this mode of operations has been stable.}
\label{fig:rates_all}       % Give a unique label
\end{figure*}

The radio emission background at the ARIANNA site is almost at the irreducible level down to \unit[50]{MHz}, as shown in Figure \ref{fig:ARIANNA-layout}. The trigger rate in the station with the upward-facing antennas is mostly dominated by the Galactic background as shown in Figure \ref{fig:rates_galaxy}. The trigger rates show an oscillation with a periodicity that corresponds to the period of the local sidereal time. Additional backgrounds are temporary signals from radio communication, which are easily vetoed by our station-level L1 trigger, which does not allow for a certain fraction of power to only be in one frequency bin. Still, one should note that the narrowband transmitters observed in the HRA are sparse, very intermittent and not necessarily strong in power. Moore's Bay seems to be one of the most radio-quiet and still accessible locations. 
The only pulsed background found at the site is correlated to heavy winds and storms and affects less than 5\% of the livetime. The HRA might register discharges within the storm or from blown snow on the surface during these distinct periods. 
As shown in Figure \ref{fig:rates_all} the stations are very sensitive to threshold settings, and further show extremely low trigger-rates. The regular stations are also affected less by the Galactic noise due to the directivity of the antennas. 

\section{Cosmic Ray detections}
The station with the upward facing antennas has been used to detect air shower signals. Using a very simple search with an increased threshold reveals a number of very interesting candidate events. Due to the large bandwidth of the HRA system and the characteristic frequency spectrum of the air shower emission, simulations show a very distinct expectation for the shape of the air shower signals. The detailed modeling of electric field from the air shower convolved with the response of the antenna, amplifier and cables predicts a time dependent waveform characterized by high to low frequency chirp, forming a well-defined pulse. The candidate events show such characteristics. 

\begin{figure*}
\centering
\includegraphics[width=0.7\textwidth]{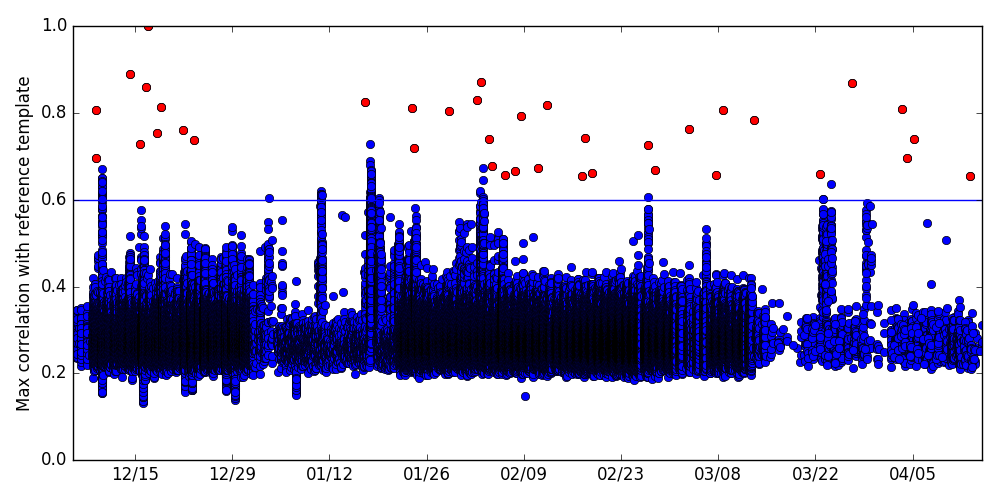}
\caption{Best correlation per event with a reference event as function of time. One event from December was chosen as reference template. It has a correlation of one. All events not in a cluster of events that have a correlation of better than 0.6 are marked in red. }
\label{fig:corr}       % Give a unique label
\end{figure*}

\begin{figure*}
\centering
\includegraphics[width=0.85\textwidth]{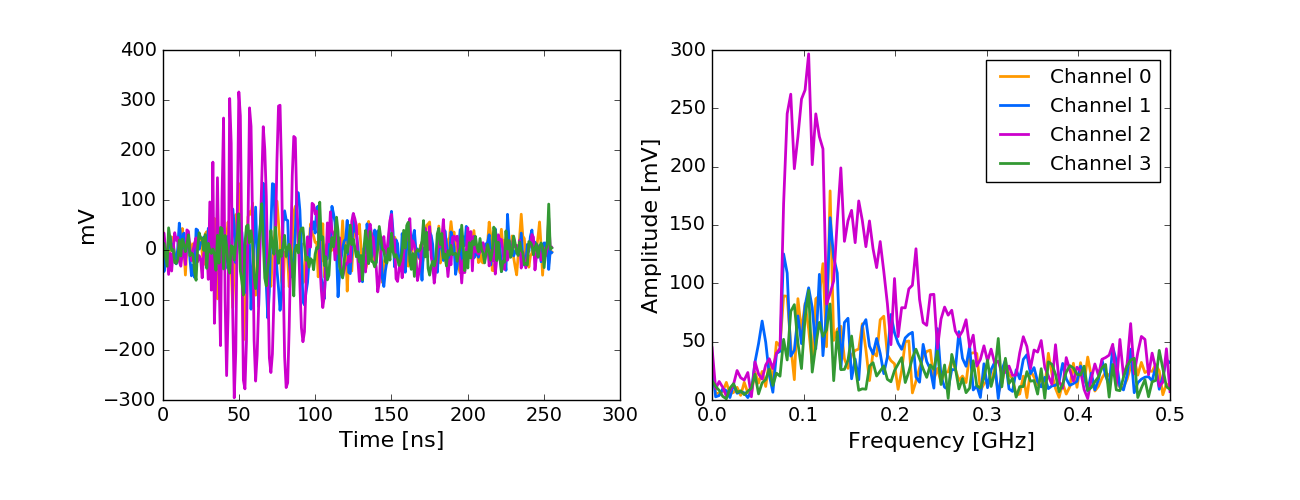}
\caption{An example of an air shower signal as measured with the HRA. Channels 2 and 3 are upward pointing channels, while 0 and 1 point downward. All signals are strongly polarized, showing only a strong amplitude in either of the perpendicular upward facing channels. The frequency spectrum of all measured air showers is dominated by antenna response and amplifier gain, however, differences can be seen especially in the high-frequency content of the pulses. }
\label{fig:CRl}       % Give a unique label
\end{figure*}

As shown earlier \cite{AstroP}, the ARIANNA Collaboration has proposed to search for neutrino signals by the use of a correlation with a template method. We now use the interesting candidates as search templates for all other events. As shown in Figure \ref{fig:corr} correlating all events with one of these template events reveals a distinct sub-set of events that show a high correlation, while most events show a very weak correlation. During periods of high winds, correlations of medium quality are observed, where some reach into a tentative signal space of above 0.6. All events marked in red are those, that do not occur in periods of high trigger-rates and show a good correlation. The events show a random occurrence in time and we observe roughly one per two days of livetime, which matches an energy threshold for cosmic rays of below \unit[$10^{18}$]{eV}. Consequently, these events are viable cosmic ray candidates. One example is shown in Figure \ref{fig:CRl}.  We will show in a forthcoming publications, how well the measured signals agree with simulated predictions and whether the number of events, in combination with livetime and efficiency, matches the known cosmic ray flux.

\section{Update on neutrino analysis}
\begin{figure*}
\centering
\includegraphics[width=0.95\textwidth]{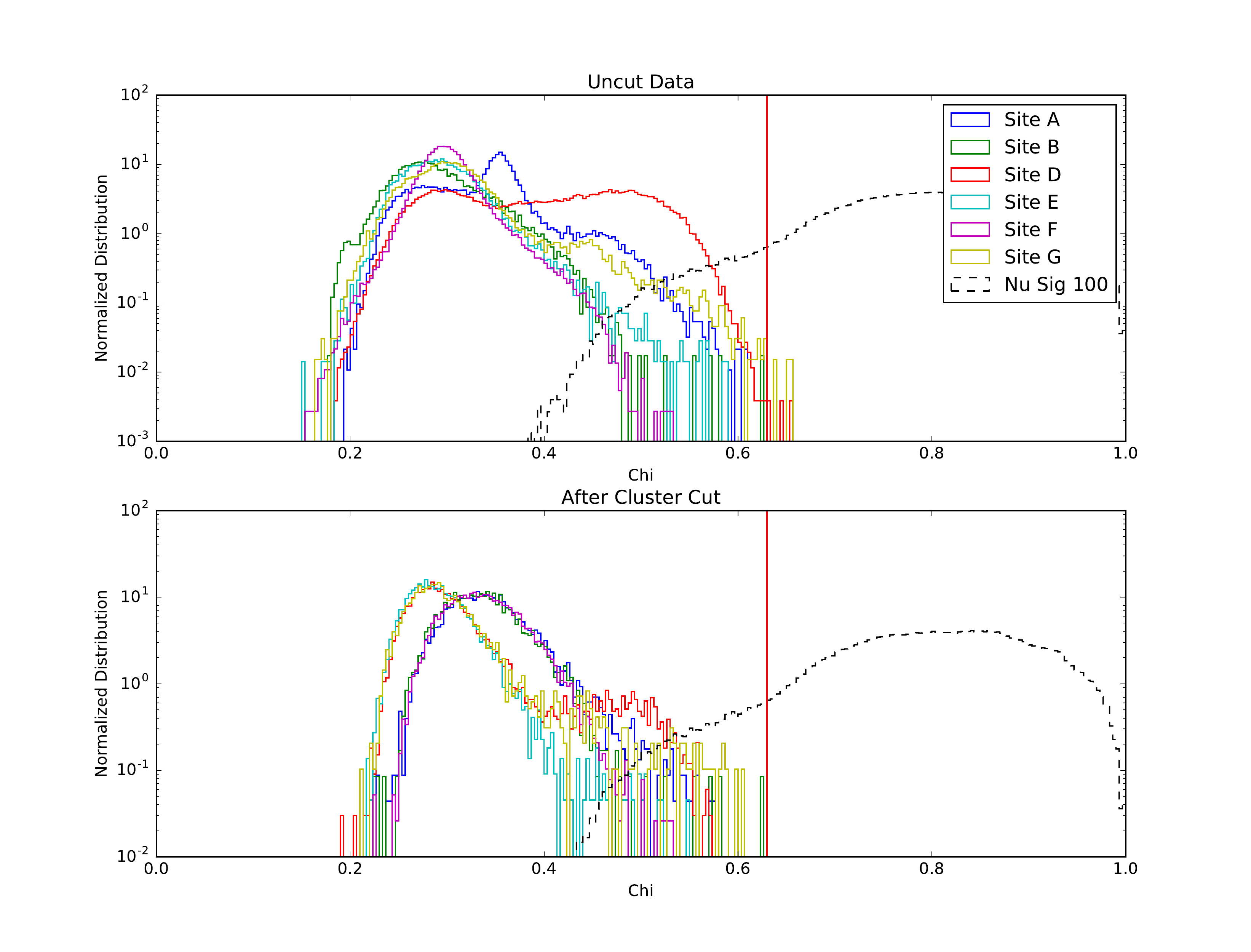}
\caption{Normalized distribution of correlation (chi) with a selected neutrino template. The two figures shows the data uncut and after the cluster cut, as explained in the text. The integral under the curves is normalized to one. The vertical line indicates a cut-off value of 0.63. The slightly differing average values of the curves are explained by two types of amplifiers.}
\label{fig:nu-distr}       % Give a unique label
\end{figure*}

\begin{figure*}
\centering
\includegraphics[width=0.5\textwidth]{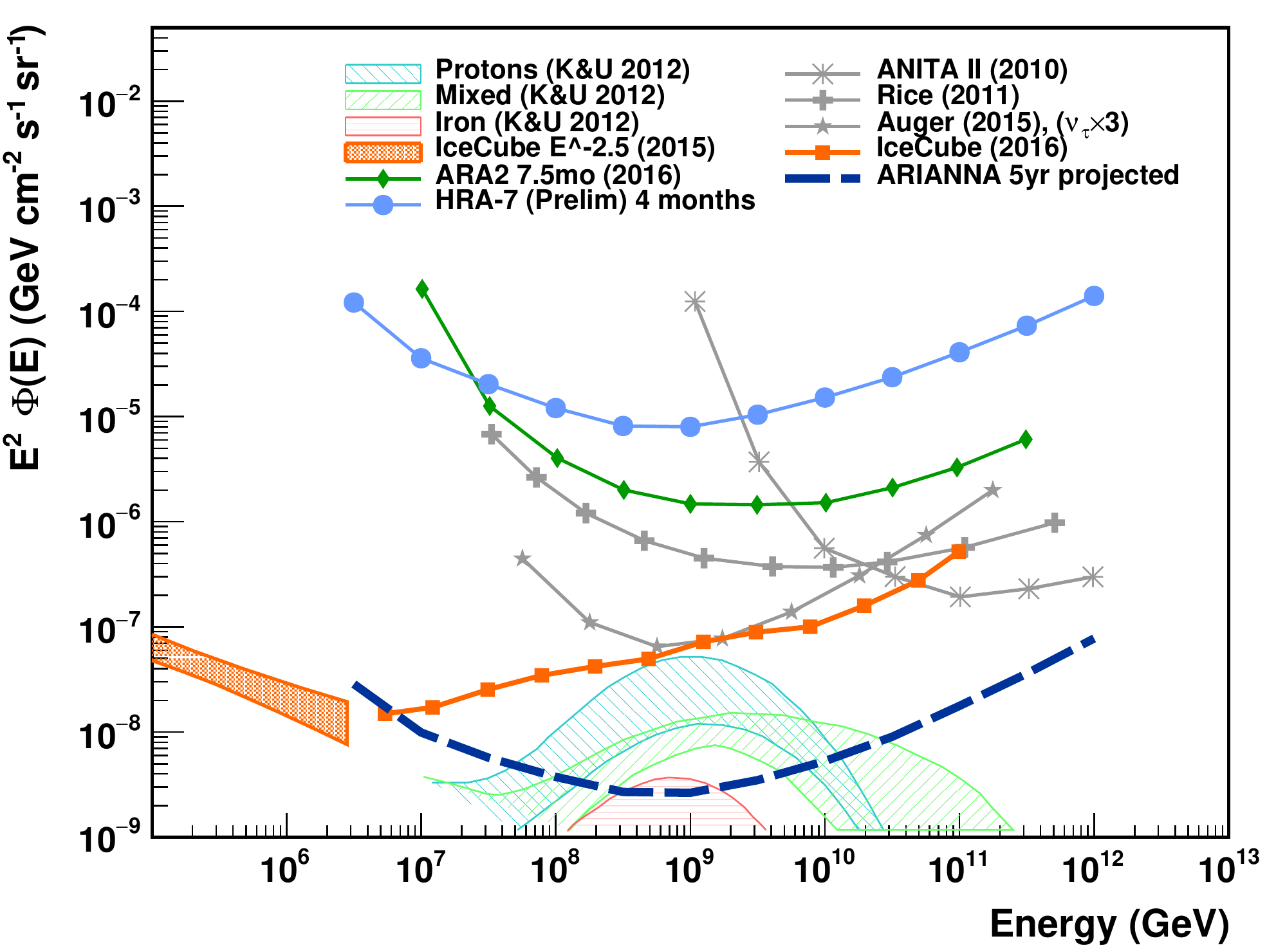}
\caption{Current limits of the HRA and projected sensitivity of ARIANNA with 1296 stations. Also shown are models of neutrino fluxes and limits of other experiments. See  \cite{ICRC,IceCube,KU} for details.}
\label{fig:lim}       % Give a unique label
\end{figure*}

The search for neutrinos follows the same approach as for cosmic rays, only that all regular stations can be used. The top panel of Figure \ref{fig:nu-distr} shows the probability density of the correlation parameter Chi \cite{AstroP} for all measured events in our data-set with a simulated neutrino signal. The bottom panel shows the distribution for the events remaining after excluding events clustered in time and those with a high single frequency contribution (L1 trigger, see above). Also shown is the expectation for Chi from the complete neutrino simulations. Most obvious is the strong peak at around 0.3 in the data, which corresponds to the correlation value of pure thermal noise, as it is also found in the forced data that is taken every 70 seconds. Some stations show a longer tail towards high values, which can be removed immediately with the cluster cut. This is explained by temporary hardware issues or increased noise at station level during storms. After cleaning, only very few events end up in the signal space. Interestingly, all of them are measured in coincidence with other stations and one has also been identified as cosmic ray in the upward-facing antennas. These are therefore the background events, which require the ARIANNA stations to have upward facing antennas for cosmic ray rejection, which will be the case for the future ARIANNA. For now, we increase the correlation threshold to exclude these events and arrive at the preliminary limits shown in Figure \ref{fig:lim} for four months of data from the season 2015/16.

\section{Conclusions and Outlook}
The HRA pilot-array is performing according to expectations with a livetime fraction of more than 0.85. The radio environment is very quiet, the stations operate reliably on battery and solar power and first cosmic rays have been detected. The neutrino analysis only reveals cosmic rays as significant background, which can easily be suppressed using upward facing antennas. The upgrade of a station with a wind-generator and with four upward facing antennas to better be able to test reconstruction algorithms on measured air shower data is planned for the next season. ARIANNA with its 1296 stations will have a a projected sensitivity to cover many of the neutrino fluxes such as shown in \cite{KU}.

\end{document}